\documentclass[twocolumn,A4]{article}
\usepackage[dvips]{graphics}
\usepackage{setspace}
\usepackage{color}
\definecolor{gold}{rgb}{0.85,0.66,0}
\definecolor{dblue}{rgb}{0,0,0.8}
\begin{document}
\onecolumn
\begin{center}
{\bf{\Large {\textcolor{gold}{Electron transport in a double quantum 
ring: Evidence of an AND gate}}}}\\
~\\
{\textcolor{dblue}{Santanu K. Maiti}}$^{1,2,*}$ \\
~\\
{\em $^1$Theoretical Condensed Matter Physics Division,
Saha Institute of Nuclear Physics, \\
1/AF, Bidhannagar, Kolkata-700 064, India \\
$^2$Department of Physics, Narasinha Dutt College,
129, Belilious Road, Howrah-711 101, India} \\
~\\
{\bf Abstract}
\end{center}
We explore AND gate response in a double quantum ring where each ring
is threaded by a magnetic flux $\phi$. The double quantum ring is 
attached symmetrically to two semi-infinite one-dimensional metallic 
electrodes and two gate voltages, namely, $V_a$ and $V_b$, are applied, 
respectively, in the lower arms of the two rings which are treated 
as two inputs of the AND gate. The system is described in the 
tight-binding framework and the calculations are done using the Green's 
function formalism. Here we numerically compute the conductance-energy 
and current-voltage characteristics as functions of the ring-to-electrode 
coupling strengths, magnetic flux and gate voltages. Our study suggests 
that, for a typical value of the magnetic flux $\phi=\phi_0/2$ 
($\phi_0=ch/e$, the elementary flux-quantum) a high output current ($1$)
(in the logical sense) appears only if both the two inputs to the gate are 
high ($1$), while if neither or only one input to the gate is high ($1$), 
a low output current ($0$) results. It clearly demonstrates the AND gate
behavior and this aspect may be utilized in designing an electronic 
logic gate. 

\vskip 1cm
\begin{flushleft}
\end{flushleft}
\vskip 4in
\noindent
{\bf ~$^*$Corresponding Author}: Santanu K. Maiti

Electronic mail: santanu.maiti@saha.ac.in

\newpage
\twocolumn

\section{Introduction}

In the present age of nanoscience and technology, the study of electron
transport through quantum confined systems has attracted a great deal of
interest, both for application and research, in the field of developing
nanoelectronics as well as spintronics. With the aid of present 
technological progress, simple looking quantum confined model systems
like, quantum rings, quantum dots, arrays of quantum dots, etc, can be
used extensively in designing nano devices, and, they are treated as the 
fundamental building blocks for future generation of nanoelectronics.
The key idea of designing nanodevices is based on the concept of quantum 
interference effect, and it is generally preserved throughout the 
sample only for much smaller sizes, while the effect disappears for 
larger systems. A normal metal mesoscopic ring is a very nice example 
where the electronic motion is confined and the transport becomes 
predominantly coherent. Using two such metallic rings, we can design 
a double quantum ring, and, here we will show how such a simple 
geometric model can be used to design an AND logic gate. To explore 
this phenomenon, we construct a bridge system where the double quantum 
ring is sandwiched between two external electrodes (Fig.~\ref{and}). 
The theoretical description of electron transfer in a bridge system 
has got much progress following the pioneering work of Aviram and 
Ratner~\cite{aviram}. Later, several excellent 
experiments~\cite{tali,reed1,reed2} have been done in several bridge 
systems to understand the basic mechanisms underlying the electron 
transport. Though in literature many theoretical~\cite{orella1,orella2,
nitzan1,nitzan2,new,muj1,muj2,walc2,walc3,cui,baer2,baer3,tagami,walc1,
baer1} as well as experimental papers~\cite{tali,reed1,reed2} on electron 
transport are available, yet lot of controversies are still present 
between the theory and experiment, and the complete knowledge of the 
conduction mechanism in this scale is not very well established even today.
 
The aim of the present work is to describe the AND gate response in a 
double quantum ring where each ring is threaded by a magnetic flux $\phi$. 
The rings are contacted symmetrically to the electrodes, and the lower
arms of the two rings are subjected to two gate voltages $V_a$ and $V_b$, 
respectively (see Fig.~\ref{and}) those are treated as the two inputs 
of the AND gate. Here we adopt a simple tight-binding model to describe 
the system and all the calculations are done numerically based on the
Green's function formalism. The AND gate 
response is illustrated by studying the conductance-energy and 
current-voltage characteristics in terms of the ring-to-electrode 
coupling strengths, magnetic flux and gate voltages. Our study reveals 
that for a typical value of the magnetic flux, $\phi=\phi_0/2$, a 
high output current ($1$) (in the logical sense) is available only if 
both the two inputs to the gate are high ($1$), while if neither or only
one input to the gate is high ($1$), a low output current ($0$) appears.
This phenomenon clearly demonstrates the AND gate response, and it may 
be utilized in designing an electronic logic gate. To the best of 
our knowledge the AND gate response in such a simple system has not 
yet been described in the literature.

The scheme of the present paper is as follow. Following the introduction 
(Section $1$), in Section $2$, we describe the model and the theoretical 
formulations for our calculations. Section $3$ presents the significant 
results, and finally, we conclude our results in Section $4$.

\section{Model and the synopsis of the theoretical background}

\begin{figure}[ht]
{\centering \resizebox*{7.7cm}{5.2cm}{\includegraphics{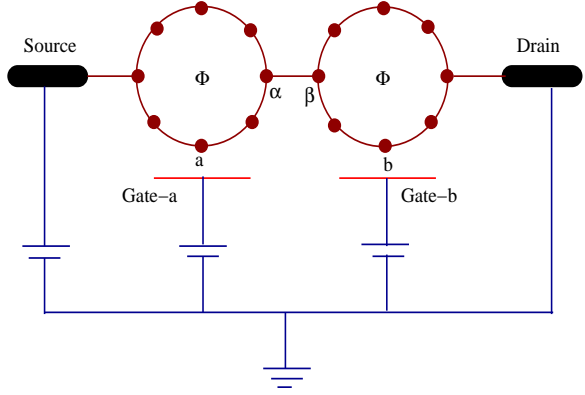}}\par}
\caption{(Color online). The scheme of connections with the batteries
for the operation of the AND gate. A double quantum ring is attached
to two semi-infinite $1$D metallic electrodes, namely, source and drain.
The gate voltages $V_a$ and $V_b$, those are variable, are applied in the
atomic sites $a$ and $b$ via the gate electrodes, gate-a and gate-b,
respectively. The source and the gate voltages are applied with respect
to the drain.}
\label{and}
\end{figure}

We start by referring to Fig.~\ref{and}. A double quantum ring, where 
each ring is threaded by a magnetic flux $\phi$, is attached symmetrically 
to two semi-infinite one-dimensional ($1$D) metallic electrodes, namely,
source and drain. Two gate electrodes, viz, gate-a and gate-b, are placed
below the lower arms of the two rings, respectively, and they are ideally
isolated from the rings. The atomic sites $a$ and $b$ in the lower arms
of the two rings are subjected to the gate voltages $V_a$ and $V_b$ via
the gate electrodes gate-a and gate-b, respectively, and they are treated 
as the two inputs of the AND gate. In the present scheme, we consider that 
the gate voltages each operate on the atomic sites nearest to the plates 
only. While, in complicated geometric models, the effect must be taken 
into account for the other dots, though the effect becomes too small.
The actual scheme of connections with the batteries for the operation 
of the AND gate is clearly presented in the figure (Fig.~\ref{and}), 
where the source and the gate voltages are applied with respect to the 
drain. 

We calculate the conductance ($g$) of the double quantum ring using the
Landauer conductance formula~\cite{datta,marc}. At much low temperatures 
and bias voltage it can be expressed as,
\begin{equation}
g=\frac{2e^2}{h} T
\label{equ1}
\end{equation}
where $T$ gives the transmission probability of an electron across
the double quantum ring. In terms of the Green's function of the
double quantum ring and its coupling to the electrodes, the 
transmission probability can be written in the form~\cite{datta,marc},
\begin{equation}
T=Tr\left[\Gamma_S G_{R}^r \Gamma_D G_{R}^a\right]
\label{equ2}
\end{equation}
where $G_{R}^r$ and $G_{R}^a$ are respectively the retarded and advanced
Green's functions of the double quantum ring including the effects of 
the electrodes. Here $\Gamma_S$ and $\Gamma_D$ describe the coupling of 
the double quantum ring to the source and drain, respectively. For the 
full system i.e., the double quantum ring, source and drain, the Green's 
function is expressed as,
\begin{equation}
G=\left(E-H\right)^{-1}
\label{equ3}
\end{equation}
where $E$ is the injecting energy of the source electron. Evaluation of 
this Green's function needs the inversion of an infinite matrix since
the full system consists of the finite double quantum ring and the two 
semi-infinite $1$D electrodes. However, the full system can be partitioned 
into sub-matrices corresponding to the individual sub-systems and the 
Green's function for the double quantum ring can be effectively written 
in the form,
\begin{equation}
G_{R}=\left(E-H_{R}-\Sigma_S-\Sigma_D\right)^{-1}
\label{equ4}
\end{equation}
where $H_{R}$ corresponds to the Hamiltonian of the double quantum ring.
Within the non-interacting picture the Hamiltonian can be expressed like,
\begin{eqnarray}
H_{R} & = & \sum_i \left(\epsilon_i + V_a \delta_{ia} + V_b \delta_{ib} 
\right) c_i^{\dagger} c_i \nonumber \\
 & + & \sum_{<ij>} t \left(c_i^{\dagger} c_j e^{i\theta}+ c_j^{\dagger} 
c_i e^{-i\theta}\right)
\label{equ5}
\end{eqnarray}
In this Hamiltonian $\epsilon_i$'s are the site energies for all the 
sites $i$ except the sites $i=a$ and $b$ where the gate voltages $V_a$ 
and $V_b$ are applied, those are variable. These gate voltages can be 
incorporated through the site energies as expressed in the above 
Hamiltonian. $c_i^{\dagger}$ ($c_i$) is the creation (annihilation) 
operator of an electron at the site $i$ and $t$ is the hopping strength 
between the nearest-neighbor sites in each ring. The hopping strength 
between the two atomic sites ($\alpha$ and $\beta$) through which 
the rings are coupled to each other is also set to $t$, for the sake 
of simplicity. $\theta=2 \pi \phi/N \phi_0$ is 
the phase factor due to the flux $\phi$ in each ring, where $N$ 
represents the total number of atomic sites (filled circles) in a 
single ring. A similar kind of tight-binding Hamiltonian is also used, 
except the phase factor $\theta$, to describe the semi-infinite $1$D 
perfect electrodes where the Hamiltonian is parametrized by constant 
on-site potential $\epsilon^{\prime}$ and nearest-neighbor hopping 
integral $t^{\prime}$. The hopping integral between the source and the
double quantum ring is $\tau_S$, while it is $\tau_D$ between the
double quantum ring and the drain. The parameters $\Sigma_S$ and 
$\Sigma_D$ in Eq.~(\ref{equ4}) represent the self-energies due to 
the coupling of the double quantum ring to the source and drain, 
respectively, where all the information of the coupling are included 
into these self-energies.

To evaluate the current ($I$), passing through the double quantum ring, 
as a function of the applied bias voltage ($V$) we use the 
relation~\cite{datta},
\begin{equation}
I(V)=\frac{e}{\pi \hbar}\int \limits_{E_F-eV/2}^{E_F+eV/2} T(E,V) dE
\label{equ8}
\end{equation}
where $E_F$ is the equilibrium Fermi energy. Here we make a realistic
assumption that the entire voltage is dropped across the ring-electrode
interfaces, and it is examined that under such an assumption the $I$-$V$
characteristics do not change their qualitative features. 

In this presentation, all the results are computed only at absolute zero
temperature. These results are also valid even for some finite (low)
temperatures, since the broadening of the energy levels of the double
quantum ring due to its coupling to the electrodes becomes much larger 
than that of the thermal broadening~\cite{datta}. On the other hand, 
at high temperature limit, all these features completely disappear. 
This is due to the fact that the phase coherence length decreases 
significantly with the rise of temperature where the contribution 
comes mainly from the scattering on phonons, and therefore, the 
quantum interference effect vanishes. For the sake of simplicity, 
we take the unit $c=e=h=1$ in our present calculations.

\section{Results and discussion}

In order to illustrate the results, let us begin our discussion
by mentioning the values of the different parameters used for the 
numerical calculations. In the double quantum ring, the on-site 
energy $\epsilon_i$ is fixed to $0$ for all the sites $i$, except 
the sites $i=a$ and $b$ where the site energies are taken as $V_a$ 
and $V_b$, respectively, and the nearest-neighbor hopping strength 
$t$ is set to $3$. While, for the 
\begin{figure}[ht]
{\centering \resizebox*{8cm}{7cm}{\includegraphics{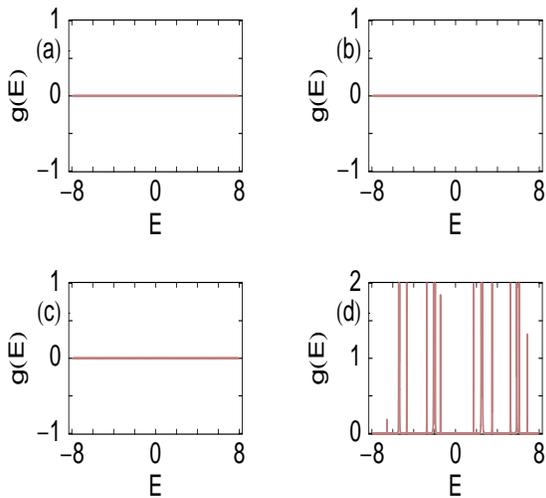}}\par}
\caption{(Color online). Conductance $g$ as a function of the energy
$E$ for a double quantum ring with $M=16$ and $\phi=0.5$ in the limit
of weak-coupling. (a) $V_a=V_b=0$, (b) $V_a=2$ and $V_b=0$, (c) $V_a=0$
and $V_b=2$ and (d) $V_a=V_b=2$.}
\label{condlow}
\end{figure}
side-attached electrodes the on-site energy ($\epsilon^{\prime}$) 
and the nearest-neighbor hopping strength ($t^{\prime}$) are chosen 
as $0$ and $4$, respectively. The Fermi energy $E_F$ is taken as $0$. 
Throughout the study, we narrate our results for the two limiting 
cases depending on the strength of the coupling of the double quantum
ring to the source and drain. Case $I$: $\tau_{S(D)} << t$. It is the 
so-called weak-coupling limit. For this regime we choose 
$\tau_S=\tau_D=0.5$. Case $II$: $\tau_{S(D)} \sim t$. This is 
the so-called strong-coupling limit. In this particular limit, we 
set the values of the parameters as $\tau_S=\tau_D=2.5$. The key 
controlling parameter for all these calculations is the magnetic 
flux $\phi$, threaded by each single ring, which is fixed at $\phi_0/2$ 
i.e., $0.5$ in our chosen unit $c=e=h=1$.

As illustrative examples, in Fig.~\ref{condlow} we plot the 
conductance-energy ($g$-$E$) characteristics for a double quantum ring
with $M=16$ ($M=2N$, the total number of atomic sites in the double 
quantum ring, since each ring contains $N$ atomic sites) in the limit 
of weak-coupling, where (a), (b), (c) and (d) 
\begin{figure}[ht]
{\centering \resizebox*{8cm}{7cm}{\includegraphics{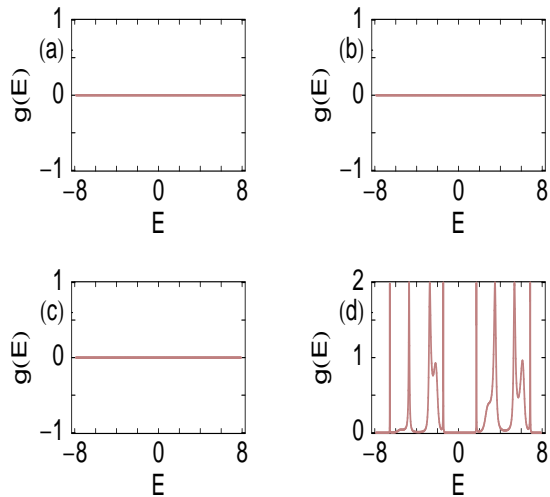}}\par}
\caption{(Color online). Conductance $g$ as a function of the energy
$E$ for a double quantum ring with $M=16$ and $\phi=0.5$ in the limit
of strong-coupling. (a) $V_a=V_b=0$, (b) $V_a=2$ and $V_b=0$,
(c) $V_a=0$ and $V_b=2$ and (d) $V_a=V_b=2$.}
\label{condhigh}
\end{figure}
correspond to the results for the four different choices of the gate 
voltages $V_a$ and $V_b$, respectively. When both the two inputs $V_a$ 
and $V_b$ are identical to zero i.e., both the inputs are low, the 
conductance $g$ becomes exactly zero for the entire energy range (see 
Fig.~\ref{condlow}(a)). A similar response is also observed for the 
other two cases where anyone of the two inputs ($V_a$ and $V_b$) to 
the gate is high and other one is low. The results are shown in 
Figs.~\ref{condlow}(b) and (c), respectively. Thus for all these three 
cases (Figs.~\ref{condlow}(a)-(c)), the double quantum ring does not allow
to pass an electron from the source to the drain. The conduction of
electron through the bridge system is allowed only when both the two
inputs to the gate are high i.e., $V_a=V_b=2$. The response is given 
in Fig.~\ref{condlow}(d), and it is observed that for some particular
energies the conductance exhibits fine resonant peaks. At the resonant 
energies, the conductance approaches the value $2$, and accordingly, 
the transmission probability $T$ goes to unity, since the relation 
$g=2T$ holds from the Landauer conductance formula (see Eq.~(\ref{equ1}) 
with $e=h=1$). All these resonant peaks are associated with the 
energy eigenvalues of the double quantum ring, and therefore, we 
can predict that the conductance spectrum manifests itself the 
electronic structure of the 
\begin{figure}[ht]
{\centering \resizebox*{8cm}{7cm}{\includegraphics{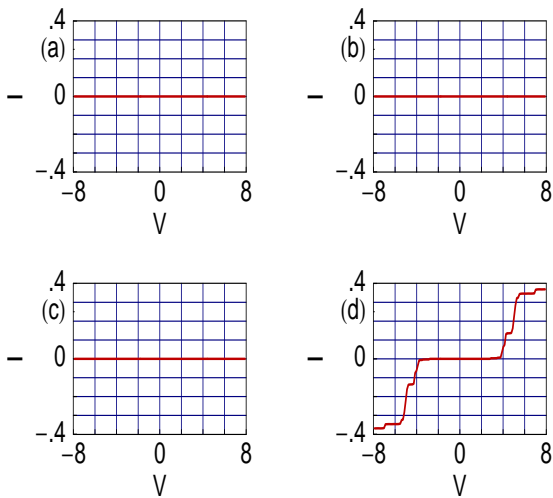}}\par}
\caption{(Color online). $I$-$V$ characteristics for a double quantum ring
with $M=16$ and $\phi=0.5$ in the weak-coupling limit. (a) $V_a=V_b=0$,
(b) $V_a=2$ and $V_b=0$, (c) $V_a=0$ and $V_b=2$ and (d) $V_a=V_b=2$.}
\label{currlow}
\end{figure}
double quantum ring. Now we try to figure out the dependences of the 
gate voltages on the electron transport in these four different cases.
The probability amplitude of getting an electron from the source to
drain across the double quantum ring depends on the combined effect 
of the quantum interferences of the electronic waves passing through 
the upper and lower arms of the two rings. For a symmetrically 
connected ring (length of the two arms of the ring are identical 
to each other) which is threaded by a magnetic flux $\phi$, the 
probability amplitude of getting an electron
across the ring becomes exactly zero ($T=0$) for the typical flux, 
$\phi=\phi_0/2$. This is due to the result of the quantum interference 
among the two waves in the two arms of the ring, which can be shown 
through few simple mathematical steps. Thus for the particular case
when both the two inputs to the gate are low ($0$), the upper and lower
arms of the two rings become exactly identical, and accordingly, the
transmission probability vanishes. The similar response i.e., the 
vanishing transmission probability, is also achieved for the two other
cases ($V_a=2$, $V_b=0$ and $V_a=0$, $V_b=2$), where the symmetry is
broken only in one ring out of these two by applying a gate voltage
either in the site $a$ or in $b$, preserving the symmetry in the other 
ring. The reason is that, when anyone of the two gates ($V_a$ and $V_b$)
is non-zero, the symmetry between the upper and lower arms is broken 
only in one ring which provides non-zero transmission probability 
across the ring. While, for the other ring where no gate voltage is
applied, the symmetry between the two arms becomes preserved which 
gives zero transmission probability. Accordingly, the combined effect
\begin{table}[ht]
\begin{center}
\caption{AND gate response in the weak-coupling limit. The current
$I$ is computed at the bias voltage $6.02$.}
\label{table1}
~\\
\begin{tabular}{|c|c|c|}
\hline \hline
Input-I ($V_a$) & Input-II ($V_b$) & Current ($I$) \\ \hline
$0$ & $0$ & $0$ \\ \hline
$2$ & $0$ & $0$ \\ \hline
$0$ & $2$ & $0$ \\ \hline
$2$ & $2$ & $0.346$ \\ \hline \hline
\end{tabular}
\end{center}
\end{table}
provides vanishing transmission probability across the bridge, as the
two rings are coupled to each other.
The non-zero value of the transmission probability is achieved only when 
the symmetries of both the two rings are identically broken. This can be 
done by applying the gate voltages in both the sites $a$ and $b$ 
of the two rings. Thus for the particular case when both the two inputs 
are high i.e., $V_a=V_b=2$, the non-zero value of the transmission 
probability appears. This feature clearly demonstrates the AND gate 
behavior. With these characteristics, we get additional one feature 
when the coupling strength of the double quantum ring to the electrodes 
increases from the low regime to the high one. In the limit of 
strong-coupling, all these resonant peaks get substantial widths compared 
to the weak-coupling limit. The results are shown in Fig.~\ref{condhigh}, 
where all the other parameters are identical to those in Fig.~\ref{condlow}. 
The contribution for the broadening of the resonant peaks in this 
strong-coupling limit appears from the imaginary parts of the 
self-energies $\Sigma_S$ and $\Sigma_D$, respectively~\cite{datta}. 
Hence by tuning the coupling strength, we can get the electron 
transmission across the double quantum ring for the wider range of 
energies and it provides an important signature in the study of 
current-voltage ($I$-$V$) characteristics.

All these features of electron transfer become much more clearly visible
by studying the $I$-$V$ characteristics. The current passing through the
double quantum ring is computed from the integration procedure of the 
transmission 
function $T$ as prescribed in Eq.~(\ref{equ8}). The transmission 
function varies exactly similar to that of the conductance spectrum, 
differ only in magnitude by the factor $2$ since the relation $g=2T$ 
holds from the Landauer conductance formula Eq.~(\ref{equ1}). As
representative examples, in Fig.~\ref{currlow} we plot the current 
$I$ as a function of the applied bias voltage $V$ for a double quantum
ring considering $M=16$ in the limit of weak-coupling, where (a), (b), (c)
and (d) represent the results for the four different cases of the two
\begin{figure}[ht]
{\centering \resizebox*{8cm}{7cm}{\includegraphics{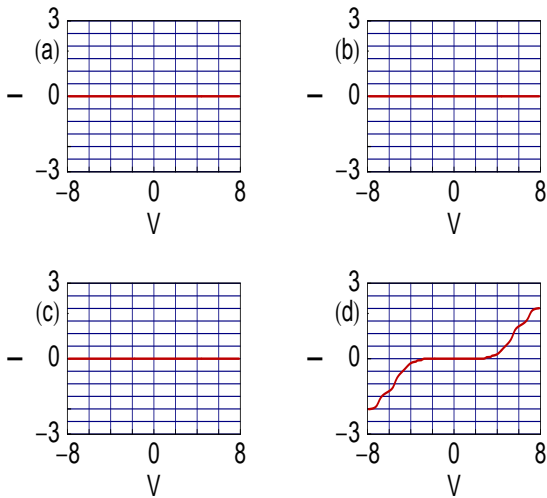}}\par}
\caption{(Color online). $I$-$V$ characteristics for a double quantum ring
with $M=16$ and $\phi=0.5$ in the strong-coupling limit. (a) $V_a=V_b=0$,
(b) $V_a=2$ and $V_b=0$, (c) $V_a=0$ and $V_b=2$ and (d) $V_a=V_b=2$.}
\label{currhigh}
\end{figure}
gate voltages $V_a$ and $V_b$. For the cases when either both the
two inputs to the gate are low ($V_a=V_b=0$), or anyone of the two 
inputs is high and other is low ($V_a=2$, $V_b=0$ or $V_a=0$, $V_b=2$), 
the current is exactly zero for the entire range of the bias voltage. 
The results are shown in Figs.~\ref{currlow}(a)-(c), and, the vanishing 
behavior of the current in these three cases can be clearly understood 
from the conductance spectra Figs.~\ref{condlow}(a)-(c), since the 
current is computed from the integration procedure of the transmission 
function $T$. The non-vanishing current amplitude is observed only for 
the typical case where both the two inputs to the gate are high i.e., 
$V_a=V_b=2$. The result is shown in Fig.~\ref{currlow}(d). 
From this figure it is observed that the current exhibits staircase-like 
structure with fine steps as a function of the applied bias voltage. 
This is due to the existence of the sharp resonant peaks in the 
conductance spectrum in the weak-coupling limit, since the current 
is computed by the integration method of the transmission function $T$. 
With the increase of the bias voltage $V$, the electrochemical potentials 
on the electrodes are shifted gradually, and finally cross one of the 
quantized energy levels of the double quantum ring. Therefore, a current 
channel 
\begin{table}[ht]
\begin{center}
\caption{AND gate response in the strong-coupling limit. The current
$I$ is computed at the bias voltage $6.02$.}
\label{table2}
~\\
\begin{tabular}{|c|c|c|}
\hline \hline
Input-I ($V_a$) & Input-II ($V_b$) & Current ($I$) \\ \hline
$0$ & $0$ & $0$ \\ \hline
$2$ & $0$ & $0$ \\ \hline
$0$ & $2$ & $0$ \\ \hline
$2$ & $2$ & $1.295$ \\ \hline \hline
\end{tabular}
\end{center}
\end{table}
is opened up which provides a jump in the $I$-$V$ characteristic curve.
Addition to these behaviors, it is also important to note that the 
non-zero value of the current appears beyond a finite value of $V$, 
the so-called threshold voltage ($V_{th}$). This $V_{th}$ can be 
controlled by tuning the size ($N$) of the two rings. From 
these $I$-$V$ characteristics, the behavior of the AND gate response
is clearly visible. To make it much clear, in Table~\ref{table1}, we 
present a quantitative estimate of the typical current amplitude, 
computed at the bias voltage $V=6.02$, in this weak-coupling limit. 
It shows $I=0.346$ only when both the two inputs to the gate are 
high ($V_a=V_b=2$), while for the other three cases when either 
$V_a=V_b=0$ or $V_a=2$, $V_b=0$ or $V_a=0$, $V_b=2$, the current 
$I$ gets the value $0$.
In the same footing, as above, here we also discuss the $I$-$V$
characteristics in the limit of strong-coupling. In this limit, the
current varies almost continuously with the applied bias voltage and
achieves much larger amplitude than the weak-coupling case 
(Fig.~\ref{currlow}) as presented in Fig.~\ref{currhigh}. The reason 
is that, in the limit of strong-coupling all the resonant peaks get 
broadened which provide larger current in the integration procedure 
of the transmission function $T$. Thus by tuning the strength of the 
ring-to-electrodes coupling, we can achieve very large current amplitude 
from the very low one for the same bias voltage $V$. All the other 
properties i.e., the dependences of the gate voltages on the $I$-$V$ 
characteristics are exactly similar to those as given in Fig.~\ref{currlow}. 
In this strong-coupling limit we also make a quantitative study for the 
typical current amplitude, given in Table~\ref{table2}, where the 
current amplitude is determined at the same bias voltage ($V=6.02$) as
earlier. The response of the output current is exactly similar to that as
given in Table~\ref{table1}. Here the non-zero value of the current gets
the value $1.295$ which is much larger compared to the weak-coupling case
which shows the value $0.346$. From these results we can clearly manifest
that a double quantum ring exhibits the AND gate response.

\section{Concluding remarks}

To summarize, we have addressed the AND gate behavior in a double quantum 
ring where each ring is threaded by a magnetic flux $\phi$. The double 
quantum ring is attached symmetrically to two semi-infinite $1$D metallic 
electrodes and two gate voltages, namely, $V_a$ and $V_b$, are applied,
respectively, in the lower arms of the two rings and they are considered 
as the two inputs of the AND gate. The full system is described by the 
tight-binding model and all the calculations are done in the Green's 
function formalism. We have numerically computed the conductance-energy 
and current-voltage characteristics as functions of the ring-electrode 
coupling strengths, magnetic flux and gate voltages. Very interestingly 
we have noticed that, for the half flux-quantum value of $\phi$ 
($\phi=\phi_0/2$), a high output current ($1$) (in the logical sense) 
appears only if both the inputs to the gate are high ($1$). On the 
other hand, if neither or only one input to the gate is high ($1$), a 
low output current ($0$) results. It clearly manifests the AND gate 
response, and, this aspect may be utilized in designing a tailor made 
electronic logic gate. In view of the potential application of this 
AND gate as a circuit element in an integrated circuit, we would like 
to mention that care should be taken during the application of the 
magnetic field in the two rings such that the other circuit elements 
of the integrated circuit are not affected by this field.

Throughout our work, we have addressed the conductance-energy and 
current-voltage characteristics for a double quantum ring with total
number of atomic sites $M=16$. In our model calculations, this typical
number ($M=16$) is chosen only for the sake of simplicity. Though the
results presented here change numerically with the ring size ($N$), but
all the basic features remain exactly invariant. To be more specific, it
is important to note that, in real situation the experimentally
achievable rings have typical diameters within the range $0.4$-$0.6$ 
$\mu$m. In such a small ring, unrealistically very high magnetic fields
are required to produce a quantum flux. To overcome this situation, 
Hod {\em et al.} have studied extensively and proposed how to construct
nanometer scale devices, based on Aharonov-Bohm interferometry, those 
can be operated in moderate magnetic fields~\cite{baer4,baer5,baer6,baer7}.

In the present paper we have done all the calculations by ignoring
the effects of the temperature, electron-electron correlation, disorder,
etc. Due to these factors, any scattering process that appears in the
arms of the rings would have influence on electronic phases, and, in
consequences can disturb the quantum interference effects. Here we
have assumed that, in our sample all these effects are too small, and
accordingly, we have neglected all these factors in this particular 
study.

The importance of this article is mainly concerned with (i) the simplicity 
of the geometry and (ii) the smallness of the size. To the best of our 
knowledge the AND gate response in such a simple low-dimensional system 
that can be operated even at finite temperatures (low) has not 
been addressed earlier in the literature.

\end{document}